\documentclass[usenatbib,longauth]{mn2e}

\usepackage{natbib}
\usepackage{longtable,lscape}
\usepackage{amsmath}
\bibpunct{(}{)}{;}{a}{}{,}
\usepackage{graphicx}

\usepackage{graphicx}
\usepackage[colorlinks=true,citecolor=blue]{hyperref}
\usepackage{color}
\usepackage{xspace}
\usepackage{dcolumn}
\usepackage{longtable}
\usepackage{lscape}
\usepackage{supertabular,booktabs}
\usepackage{url}
\usepackage{afterpage}

\newcommand{\MJup}{M$_{\mathrm{Jup}}$\xspace}

\newcommand{\MSun}{M$_{\odot}$\xspace}
\newcommand{\LSun}{L$_{\odot}$\xspace}

\newcommand{\MEarth}{M$_{\oplus}$\xspace}

\newcommand{\mic}{$\mu$m\xspace}
\newcommand{\as}{\hbox{$^{\prime\prime}$}\xspace}

\title[]{Determining mass limits around HD\,163296 through SPHERE direct
  imaging data}
\author[D. Mesa et al.]{D. Mesa$^{1,2}$\thanks{E-mail:
    dino.mesa@inaf.it (AVR)}, M. Langlois$^{3,4}$, A. Garufi$^{5}$, R. Gratton$^{1}$, S. Desidera$^{1}$, V. D'Orazi$^{1}$, \newauthor
  O. Flasseur$^{6}$, M. Barbieri$^{2}$, M. Benisty$^{7,8}$, T. Henning$^{9}$, R. Ligi$^{10}$, E. Sissa$^{1}$, A. Vigan$^{4}$, \newauthor
  A. Zurlo$^{11,12,4}$, 
  A. Boccaletti$^{13}$, M. Bonnefoy$^{7}$, F. Cantalloube$^{8}$, G. Chauvin$^{7}$, \newauthor A. Cheetham$^{14}$, V. De Caprio$^{15}$,
  P. Delorme$^{8}$, M. Feldt$^{9}$, T. Fusco$^{16,4}$, L. Gluck$^{8}$, \newauthor J. Hagelberg$^{17}$, A.-M. Lagrange$^{8}$, C. Lazzoni$^{1}$, F. Madec$^{4}$, A.-L. Maire$^{9,18}$, F. Menard$^{8}$, \newauthor M. Meyer$^{19,17}$, J. Ramos$^{9}$, E.L. Rickman$^{14}$, D. Rouan$^{13}$, T. Schmidt$^{20,13}$, \newauthor G. Van der Plas$^{7}$\newauthor
  \\ 
  $^{1}$INAF-Osservatorio Astronomico di Padova, Vicolo dell'Osservatorio 5, Padova, Italy, 35122-I\\
  $^{2}$INCT, Universidad De Atacama, calle Copayapu 485, Copiap\'{o}, Atacama, Chile\\
  $^{3}$Univ. Lyon, Univ. Lyon 1, ENS de Lyon, CNRS, CRAL UMR 5574, 69230 Saint-Genis-Laval, France \\
  $^{4}$Aix Marseille Univ., CNRS, CNES, LAM, Marseille, France \\
  $^{5}$INAF, Osservatorio Astrofisico di Arcetri, Largo Enrico Fermi 5, 50125 Firenze, Italy \\
  $^{6}$Universit\'{e} de Lyon, UJM-Saint-Etienne, CNRS, Institut d'Optique Graduate School, Laboratoire Hubert Curien UMR 5516, F-42023, Saint-Etienne, France\\
  $^{7}$Unidad Mixta Internacional Franco-Chilena de Astronomía (CNRS, UMI 3386), Departamento de Astronom\'{i}a, Universidad de Chile, \\   Camino El Observatorio 1515, Las Condes, Santiago, Chile \\
  $^{8}$Univ. Grenoble Alpes, CNRS, IPAG, 38000 Grenoble, France\\
  $^{9}$Max-Planck-Institut f\"ur Astronomie, K\"onigstuhl 17, 69117, Heidelberg, Germany \\
  $^{10}$INAF-Osservatorio Astronomico di Brera, Via E. Bianchi 46, I-23807, Merate, Italy \\
  $^{11}$Nucleo de Astronomia, Facultad de Ingenieria y Ciencias, Universidad Diego Portales, Av. Ejercito 441, Santiago, Chile \\
   $^{12}$Escuela de Ingenieria Industrial, Facultad de Ingenieria y Ciencias, Universidad Diego Portales, Av. Ejercito 441, Santiago, Chile \\
   $^{13}$LESIA, Observatoire de Paris, Universit\'{e} PSL, CNRS, Sorbonne
  Universit\'{e}, Univ. Paris Diderot, Sorbonne Paris Cit\'{e}, 5 place Jules
  \\ Janssen, F-92195 Meudon, France \\
  $^{14}$Geneva Observatory, University of Geneva, Chemin des Maillettes 51, 1290 Versoix, Switzerland \\
  $^{15}$INAF - Osservatorio Astronomico di Capodimonte, Salita Moiariello 16, 80131 Napoli, Italy\\
  $^{16}$DOTA, ONERA, Universit\'{e} Paris Saclay, F-91123, Palaiseau France \\
  $^{17}$Institute for Particle Physics and Astrophysics, ETH Zurich, Wolfgang-Pauli-Strasse 27, 8093 Zurich, Switzerland \\
  $^{18}$STAR Institute, Universit\'{e} de Li\'{e}ge, All\'{e}e du Six Ao\^{u}t 19c, B-4000, Li\'{e}ge, Belgium \\
  $^{19}$Department of Astronomy, University of Michigan, 1085 S. University Ave, Ann Arbor, MI 48109-1107, USA \\
  $^{20}$Hamburger Sternwarte, Gojenbergsweg 112, 21029 Hamburg, Germany \\
}
\begin{document}

\date{Accepted . Received ; in original form }

\pagerange{\pageref{firstpage}--\pageref{lastpage}} \pubyear{}

\maketitle

\label{firstpage}

\begin{abstract}
  HD\,163296 is a Herbig Ae/Be star known to host a protoplanetary disk
  with a ringed structure. To explain the disk features, previous
  works
  proposed the presence of planets embedded into the disk. We have observed
  HD\,163296 with the near-infrared (NIR) branch of SPHERE composed by
  IRDIS and IFS with the aim to put tight constraints on the presence of
  substellar companions around this star. Despite the low rotation of the
  field of view during our observation we were able to put upper mass limits
  of few \MJup around this object. These limits do not allow to give any
  definitive conclusion about the planets proposed through the disk
  characteristics. On the other hand, our results seem to exclude the presence
  of the only candidate proposed until now using direct imaging in the NIR even
  if some caution has to be taken considered the different wavelength bands
  of the two observations.
\end{abstract}


\section{Introduction}
\label{s:intro}

The most promising environments to search for in-formation planetary
systems are protoplanetary disks around very young stars
\citep[see, e.g., ][]{2012ApJ...756..133C,2014A&A...565A..15M}. These systems
can be probed both with high-contrast imaging in the near-infrared through
instruments like SPHERE \citep{2019arXiv190204080B}, GPI
\citep{2014PNAS..11112661M}, Keck/NIRC2 \citep{2017AJ....153...44M} and
  CHARIS at Subaru Telescope \citep{2015SPIE.9605E..1CG} and at sub-millimeter
wavelengths with instruments
like ALMA. One noteworthy example of the first case is the recently discovered
planet around the disk hosting star PDS\,70
\citep{2018A&A...617A..44K,2018A&A...617L...2M}. On the other hand, in recent
years an increasing number of protoplanetary disk with gaps and rings have
been imaged through ALMA \citep[e.g., ][]{2015ApJ...808L...3A,2016ApJ...820L..40A,2017ApJ...840...23L,2018A&A...610A..24F,2018ApJ...859...21A,2018ApJ...869L..43H,2018ApJ...859...32P}. 
One of the most promising model to explain these structures implies that they
are due to the interactions between the disk and planetary mass objects
\citep[e.g., ][]{1999ApJ...514..344B,2016ApJ...818...76J}. However, plenty
of alternative models have been proposed to explain these structures including
dust accumulations at the snowlines \citep[e.g., ][]{2015ApJ...806L...7Z},
zonal flows \citep[e.g., ][]{2017A&A...600A..75B} or secular gravitational
instability \citep[e.g., ][]{2014ApJ...794...55T}. Clearly, having the
possibility to directly image the foreseen planetary companions into these
disks or, alternatively, to put tight limits on the masses of these objects
could help to disentangle between the proposed models. \par
HD\,163296 (HIP\,87819) is an A1V spectral type \citep{2001A&A...378..116M}
Herbig Ae/Be star at a distance of $101.5\pm1.2$~pc from the Sun
\citep{2018A&A...616A...1G}. Recently, its stellar parameters were 
defined by \citet{2018ApJ...869..164S} fitting its H- and K-band flux with
the {\it PARSEC} models \citep{2012MNRAS.427..127B} and finding an age of
10.4~Myr and a mass of 1.9~\MSun.
We will assume these parameters in this work. It is
important, however, to note that there is a discrepancy between the age we are
assuming and the evolutionary stage of the disk around this star as deduced by
the observations (see below). From this point of view, the previous age
determination around 4-5~Myr
\citep[e.g., ][]{1998A&A...330..145V,2009A&A...495..901M} would be
in a better agreement with the disk evolutionary stage. In any case, with an
estimated $T_{eff}$ of $\sim$9200~K, we would need a luminosity about two times
the predicted 16~\LSun to be able to fit the evolutive track of a 4-5~Myr
star. In addition, the previous determinations of the age of the star were
obtained assuming a distance of $\sim$122~pc and not the more recent value
given above. \par
The presence of dust associated to this star was first demonstrated through
observations at millimeter wavelengths
\citep[see e.g., ][]{1997ApJ...490..792M}. Observations in the infrared (IR)
proved the presence of warm gas and silicate in the disk
\citep[e.g., ][]{1999ApJ...510..408S} while observations in the visible
allowed \citet{2000ApJ...544..895G} to define a radius as $\sim$500~au. The
mass of the disk was estimated between 0.01 and 0.15~\MSun
\citep{2007A&A...469..213I,2012A&A...538A..20T}. A first
detection of the presence of a ring structure in this disk was obtained by
\citet{2014A&A...568A..40G,2017A&A...603A..21G} using polarized near-infrared
(NIR) data taken with NACO at the VLT. \par
More recently, \citet{2016PhRvL.117y1101I} revealed, using ALMA data taken with
a resolution of 20~au, the presence of three dark concentric gaps at 45, 87
and 140~au and of three bright rings at 68, 100 and 160~au from the star. Like
for other works on this target, they used a distance of 122~pc for the system
\citep{1997A&A...324L..33V} instead of the more recent value cited above. Here
and for all the works that used the old distance that are cited in this paper,
we have then updated the value of the separations using the
updated value of the distance. To explain the ring structure, they proposed
that the two external gaps were due to the presence of two planets, both of
them with a mass of $\sim$0.3~\MJup. On the other hand, they were not able to
explain the inner gap with the presence of a single planet. It has to be
explained by a different physical process or by the presence of more than one
planet with a Saturn-like mass. However, using the same data and comparing them
with 2D-hydrodynamic simulations, \citet{2018ApJ...857...87L} were able to
explain all of the gaps with the presence of planets with masses of 0.46, 0.46
and 0.58~\MJup at a separation of 48, 86 and 131~au respectively. To reach this
goal they have to assume a disk with a viscosity variable from less than
$10^{-4}$ in the inner disk to $\sim7.5\times10^{-3}$ in the outer disk. On the
other hand, \citet{2018ApJ...867L..14V} demonstrated that it is possible to
explain the gaps through models with grain growth at radii corresponding to
the snowlines of molecules like e.g. $CH_4$ and CO. Afterwards, the presence
of two planets was proposed by \citet{2018ApJ...860L..12T} through a new method
to measure the rotation curve of the CO into the disk that allowed to determine
substantial deviations from the Keplerian velocity. The two proposed planets
are at a separation of 83 and 137 au from the central star and have a mass of
1 and 1.3~\MJup, respectively. A further planetary mass companion was proposed
by \citet{2018ApJ...860L..13P} that found a localized deviation from the
Keplerian velocity of the molecular gas in the protoplanetary disk.
In this latter case, they
were also able to propose not only a separation for the planet, but also
its approximate position. Indeed, the proposed planet has a
separation of $2.3\pm0.2$\as (corresponding to a distance of around 260~au)
from the central star and a position angle of $-3^{\circ}\pm5^{\circ}$. The mass
of this object should be $\sim$2~\MJup. Finally,
\citet{2018ApJ...866..110D} proposed, using two-fluids hydrodynamics
simulations, that a single planet with a mass of 65~\MEarth at a separation
of 108~au can account for the ring structure of the disk if it has a very low
viscosity (less than $10^{-4}$). \par
HD\,163296 was also observed in the NIR with GPI \citep{2017ApJ...838...20M}
in J-band polarized light, with SPHERE using the polarized mode of
IRDIS \citep{2014SPIE.9147E..1RL} both in J- and H-band
\citep{2018A&A...614A..24M}  and with HiCIAO and CHARIS at Subaru both
  in polarized light and high contrast spectroscopy
  \citep{2019ApJ...875...38R}. In
all these cases, they were able to observe only
the ring at $\sim$67~au from the star.
\par
The star was also observed in high-contrast imaging with Keck/NIRC2 by
\citet{2018MNRAS.479.1505G} in L' spectral band. They were able to put a mass
limit of 8-15~\MJup, 4.5-6.5~\MJup and 2.5-4~\MJup at the position of the
three gaps detected by \citet{2016PhRvL.117y1101I}. Furthermore, they identified
a point source at $\sim$0.5\as and at a position
angle of $\sim30^{\circ}$. This would correspond to a distance of 67~au from
the star near the inner edge of the second gap in the disk while they
proposed for this object a mass of $\sim$6~\MJup. Regarding the disk,
also in this case only the inner bright ring of the disk was partially
imaged. \par
Finally, a new ALMA observation in the context of the DSHARP
\citep{2018ApJ...869L..41A} project was taken at a resolution
of $\sim$4~au by \citet{2018ApJ...869L..49I}. This allowed to confirm the ring
structure found in the previous observation. Moreover a new gap-bright ring
combination was identified at $\sim$10~au from the star together with a
substructure in the first ring. \par
The large disk mass together with the relatively old stellar age is a further
hint of the presence of embedded planets as it supports the existence of
dust traps generated by the presence of companions preventing the accretion
of material on the star for a long timescale \citep{2018A&A...620A..94G}. \par
We have used SPHERE to observe HD\,163296 in high-contrast imaging with the
aim to disentangle between the proposed models explaining its disk structure
and to put limits on the masses of substellar objects around this star.
In this paper we give the results obtained from these observations
that were obtained in the context of the SHINE (SPHERE High-contrast imaging
survey for Exoplanets) survey \citep{2017sf2a.conf..331C}. In
Section~\ref{s:data} we present the dataset and detail the data reduction
method while in Section~\ref{s:result} we display the results. Finally, in
Section~\ref{s:conclusion} we discuss them and give the conclusions.

\section[]{Observations and data reduction}
\label{s:data}

HD\,163296 was observed during the nights of 2017-09-29 and 2018-05-06 with
SPHERE operating in the IRDIFS\_EXT mode. In this mode, IRDIS
\citep[infrared dual-band imager and spectrograph; ][]{2008SPIE.7014E..3LD}
operates in dual-band imaging \citep{Vig10} configuration with the K1-K2
filters (K1=2.110~\mic; K2=2.251~\mic) while IFS
\citep[integral field spectrograph; ][]{Cl08} works in the Y-H spectral bands
between 0.95 and 1.65~\mic. In the first epoch the total exposure time
  was of 1536~s with IRDIS and of 1728~s with IFS while in the second epoch
it was of 4608~s with both instruments. 
The weather conditions of the two epochs are detailed in Table~\ref{t:obs}
and were generally good. However, the main limitation to the contrast that
can be obtained from these observations is the low rotation of the field of
view (FOV) especially for what concerns the first epoch, as can be seen in
Table~\ref{t:obs}. This is due to the fact that the declination of HD\,163296
is very near to the Paranal Observatory latitude preventing from observing it
during the passage of the star at meridian because of VLT pointing restrictions
at less than 3 degrees from the zenith. This is a severe limit to the
total rotation of the FOV. \par
\begin{table*}
 \centering
 \begin{minipage}{140mm}
   \caption{Characteristics of the SPHERE observations of HD\,163296. In column
     2 we list the observing mode used (see the text for more details) and in
     column 3 we report the coronagraph used. In columns
     4 and 5 we list the number of datacubes, the number of frames for each
     datacube and the exposure time, given in s, for each frame for the
     IRDIS and IFS data, respectively. In column 6 we list the total rotation
     of the FOV during the observation, while in columns 7, 8 and 9 we report
     the median values of seeing, coherence time and wind speed during the
     observations. \label{t:obs}}
  \begin{tabular}{c c c c c c c c c}
  \hline
    Date & Obs. mode & Coronagraph &  Obs. IRDIS  & OBs IFS & R.A. ($^{\circ}$) & S (\as) & $\tau_{0}$ (ms) & wind (m/s)\\
         \hline
         2017-09-29  & IRDIFS\_EXT & N\_ALC\_YJH\_S  &  4$\times$12;32  &  18$\times$12;8 &   1.33  &  0.68  & 5.4  &  3.38\\
         2018-05-06  & IRDIFS\_EXT & N\_ALC\_Ks  &  9$\times$16;32  &  3$\times$16;96 &  16.02  &  0.67  & 3.7   &  9.07 \\
\hline
\end{tabular}
\end{minipage}
\end{table*}
Both IRDIS and IFS data were reduced using the SPHERE data reduction and
handling \citep[DRH; ][]{2008SPIE.7019E..39P} pipeline exploiting the SPHERE
data center \citep{2017sf2a.conf..347D} interface. We also used, to implement
the speckle subtraction procedures, the SpeCal tool \citep{2018A&A...615A..92G}
appositely developed for SPHERE data reduction.
IFS data reduction was performed using the procedure described by
\citet{zurlo2014} and by \citet{mesa2015} to create calibrated datacubes
composed of 39 frames at different wavelengths on which we applied the
principal components analysis
\citep[PCA; e.g.][]{2012ApJ...755L..28S,2015A&C....10..107A} to reduce the
speckle noise. This algorithm allowed us to implement at the same time both
angular differential imaging \citep[ADI; ][]{2006ApJ...641..556M} and
spectral differential imaging \citep[SDI; ][]{1999PASP..111..587R}. The
self-subtraction was appropriately taken into account by
injecting in the data fake planets at different separations.
IRDIS data were reduced following the procedure described by
\citet{2016A&A...587A..57Z} and applying the PCA algorithm for the reduction
of the speckle noise. For all the dataset the contrast was calculated
following the procedure described by \citet{mesa2015} taking
into account the small sample statistics as devised in
\citet{2014ApJ...792...97M}. Despite the limitations of the adopted
  method, showed by recent works
  \citep{2018AJ....155...19J,2017AJ....154...73R}, we found that the approach
  that we adopted is able to provide reliable limits for the present case.\par
An alternative data reduction have been performed using PACO 
\citep{2018A&A...618A.138F} for IRDIS and PACO-ASDI (Flasseur et al., in prep.)
for IFS. Contrary to existing approaches, this method models the background statistics to locally capture the spatial (PACO) and spectral (PACO-ASDI)
correlations at the scale of a patch of a few tens of pixels. Since PACO locally learns the background fluctuations, the aberrant data or the larger stellar leakages can also be learned locally as typical background fluctuations and are not interpreted in the detection stage as the signature of an exoplanet. The method produces both stationary and statistically grounded detection maps, as well as the false alarm rate and the probability of detection, that have been proven to be robust by fake planet injections. The detection maps are robust to defective pixels and other aberrant data points arising during the SPHERE observations or data pre-processing pipeline. The patches considered in the PACO algorithm define the characteristic size of the areas in which the statistics of the background fluctuations are modelled. Their size obeys a trade-off: on the one hand, the larger the patches, the more energy from the source is contained in the patches which improves the signal-to-noise ratio; on the other hand, learning the covariance of larger patches requires more temporal diversity. In practice, since the sources to be detected are faint compared to the level of stellar speckles and their temporal fluctuations, the optimal patch size corresponding to twice the off-axis PSF full width at half maximum (FWHM) is used (leading to patch radii of 4 and 5 pixels for K1 and K2 filters respectively) to produce the contrast limits (Fig. \ref{f:contrast}). However, given that this method is new, we will use in the following the PACO contrast values as a lower limit while the PCA contrast will be used as conservative values for the contrast. 
\section{Results}
\label{s:result}

\begin{figure*}
\centering
\includegraphics[width=\textwidth]{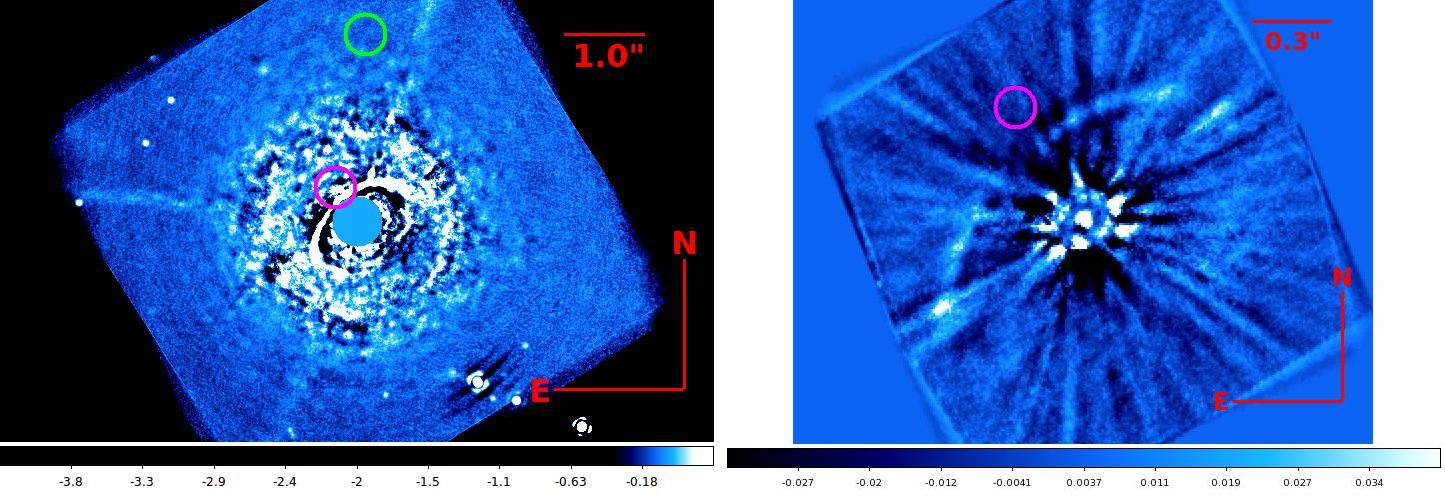}
\caption{Final images obtained from the 2018-05-06 observations. In both
  cases the values are expressed in counts. {\it Left:}
  IRDIS final image obtained applying the PCA algorithm and
  subtracting 1 mode from the original data. The green circle is to enligth
  the zone around the planet proposed by \citet{2018ApJ...860L..13P},
  the magenta circle enlight the zone of the planet proposed by
  \citet{2018MNRAS.479.1505G}. The mask covering the star has a radius
    of 0.37\as. {\it Right:} IFS final image obtained using the
  PCA algorithm subtracting 1 mode from the original data. As for the IRDIS
  image, the magenta circle identify the zone around the planet proposed by
  \citet{2018MNRAS.479.1505G}.}
\label{f:finalimage}
\end{figure*}

\begin{figure*}
\centering
\includegraphics[width=\textwidth]{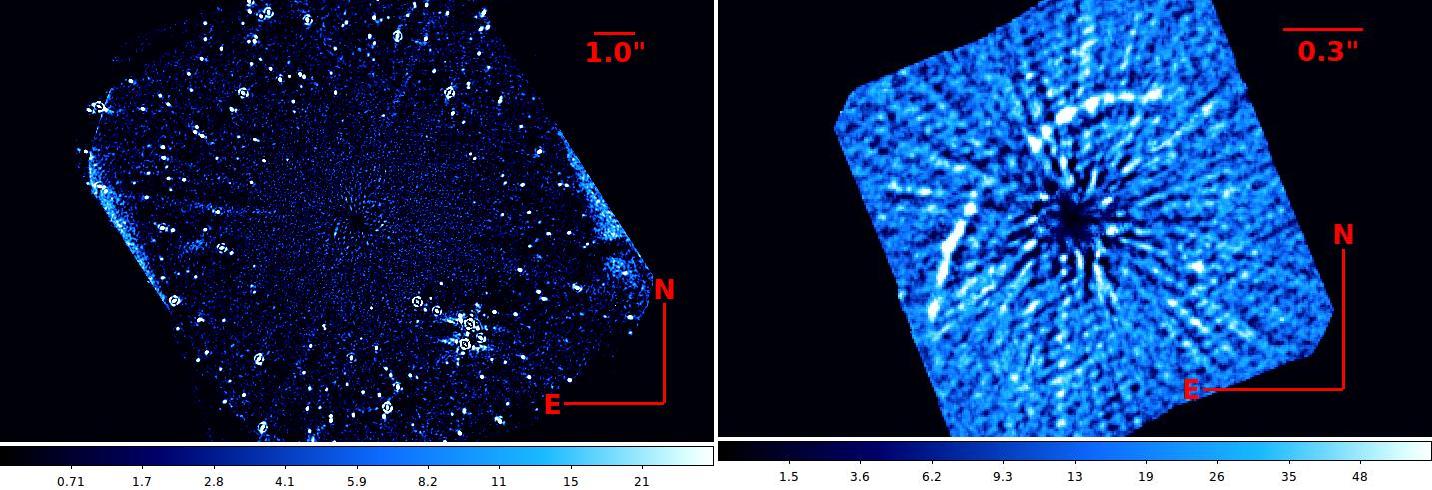}
\caption{GLR maps obtained using PACO from the 2018-05-06 observations. In the
  {\it left} panel we display the IRDIS image while in the {\it right} panel we
  display the IFS image.}
\label{f:finalimagepaco}
\end{figure*}

In Figure~\ref{f:finalimage} we display the final images that we obtain
from the IRDIS (left panel) and IFS (right panel) data using the PCA-based
data reduction while in Figure~\ref{f:finalimagepaco} we display the generalized
likelihood ratio \citep[GLR; ][]{2018A&A...618A.138F} map obtained from PACO
and PACO-ASDI for IRDIS and IFS, respectively.
Within the PACO framework, GLR is defined by
$\text{GLR} = \sum_{\lambda=1}^L \text{max}(\text{SNR}_{\lambda},0)^2$.
While it is not statistically grounded when $L \neq 1$ (as it the case for
the IFS image), it is used here as a simple combination of the available
spectral channels to emphasize structures at weak level of contrast.

\subsection[]{Disk detection}
\label{s:diskres}

Like for previous NIR observations we are able to detect only the first ring
of the disk. As pointed out by \citet{2018A&A...614A..24M}, this is probably
due to the lack of small dust grain on the surface of the outer disk.

The signal to noise
ratio (SNR) of the detection is however low both with IRDIS (with a value of
$\sim$7.5 in the brightest part of the disk and median values below 3) and IFS
(with values around or below 2) and, in particular in the IFS case, it appears
incomplete.
Despite this is not the main goal of our observations, we derived
the main parameters of the disk following a procedure similar
to that devised in \citet{2019AJ....157...39G} using the model Zodipic
\citep{2012ascl.soft02002K}. This procedure is aimed to maximize
the cross-correlation between synthetic disks obtained using the Zodipic model
with different parameters and our data.
To this aim we used the second epoch IRDIS data that allowed the best
  imaging of the disk between our data.
At the end of this procedure we found for the bright ring an inclination
of $49.3\pm2.0^{\circ}$, a position angle of $133.6\pm2.0^{\circ}$ and radius
of $\sim63\pm3$~au. The values of the offset were of $-16.6\pm10.0$~mas
  in X and of $-0.5\pm10.0$~mas in Y. These values are similar to those
obtained from previous work aimed to study the disk
\citep[e.g., ][]{2018A&A...614A..24M,2018ApJ...869L..49I}, but of course the
results are plagued by the low SNR obtained for the disk from our data.

\subsection{Detection limits}
\label{s:detlim}

\begin{figure}
\centering
\includegraphics[width=\columnwidth]{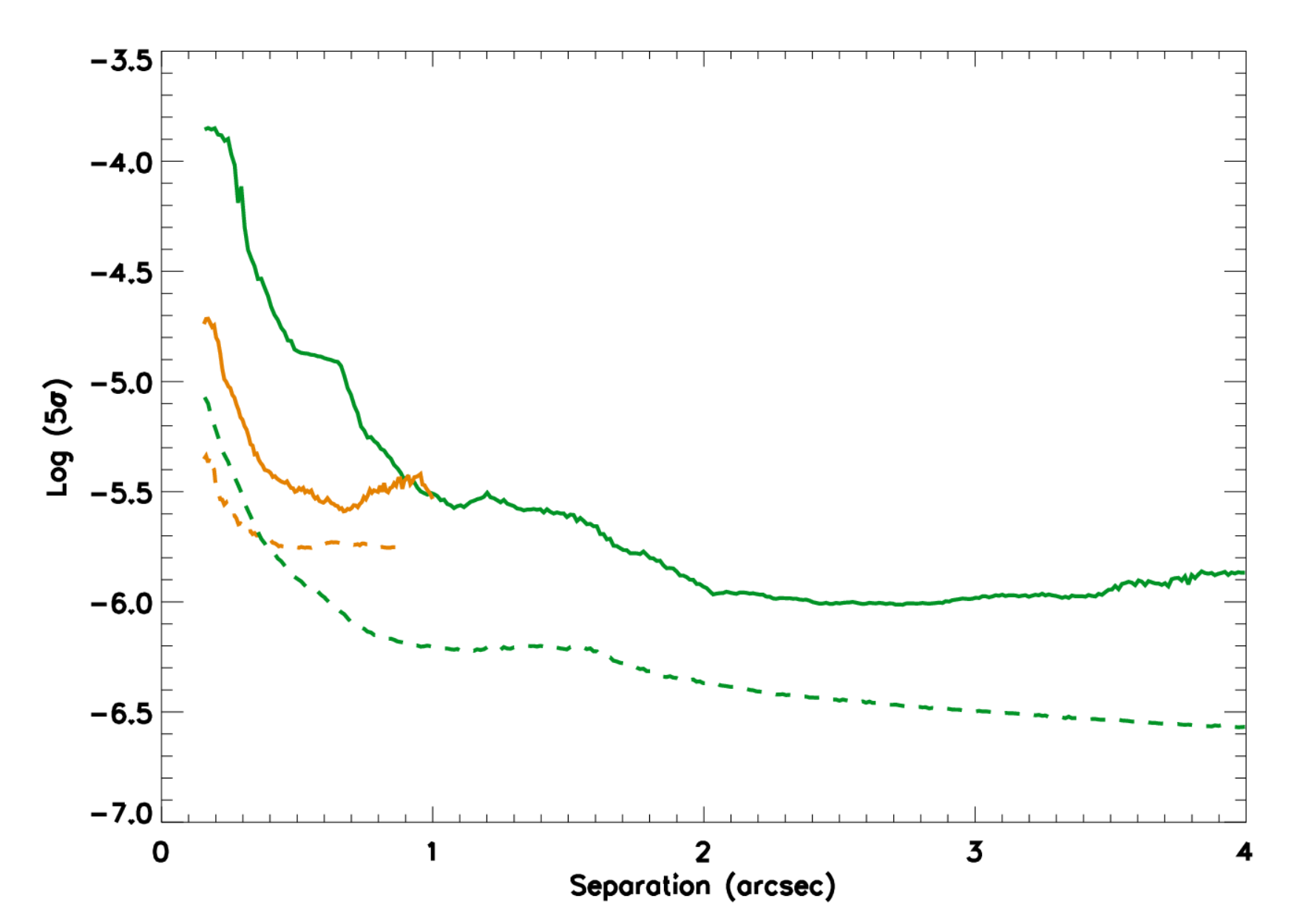}
\caption{5$\sigma$ contrast vs separation expressed in arcsec obtained for
  IRDIS (solid green line) and for IFS (solid orange line) using the PCA
  algorithm and subtracting 1 mode from the original data as done for the
    images shown in Figure~\ref{f:finalimage}. The dashed lines represent
  the contrast obtained using PACO
  using patch radii of 4 and 5 pixels for K1 and K2 filters respectively.}
\label{f:contrast}
\end{figure}

Deep high-contrast imaging data allows to put much tighter
constraints on the mass of possible substellar companion around HD\,163296
than the polarized data. In Figure~\ref{f:contrast} we display the contrast
versus the separation both for IRDIS (green line) and IFS (orange line) using
the data from the second epoch, when we were able to obtain a deeper contrast,
applying the PCA algorithm. We also display, using dashed lines of the same
colors of the PCA plots, the contrast obtained using the PACO algorithm both
for IRDIS and IFS.  To take into account the inclination of the disk, we have
deprojected the separations on our images adopting an inclination of
$46.7^{\circ}$ and a position angle of $133.3^{\circ}$ as found by
\citet{2018ApJ...869L..49I}. Despite the low rotation of the FOV, we are able
to obtain a contrast better than $10^{-5}$ at separations of few tenths of
arcsec with IFS while IRDIS allows to obtain a contrast of the order of
$10^{-6}$ at separation larger than 2\as. PACO allows to obtain a gain of
around three times for the IRDIS case while the gain is less
important in the IFS case. \par
For what concerns the planets proposed in previous works to explain the disk
structure, in Figure~\ref{f:finalimage} we have highlighted, both in IRDIS and
IFS image, with a magenta circle the region around the position of the
6~\MJup planet proposed by \citet{2018MNRAS.479.1505G} through direct imaging.
In both cases, we are not able to find any evidence of the proposed companion.
Moreover, in the IRDIS image we also display a green circle to enlight the
region around the 2~\MJup planet proposed by \citet{2018ApJ...860L..13P}.
Also in this case, the proposed planet is not visible in our data. \par
\begin{figure}
\centering
\includegraphics[width=\columnwidth]{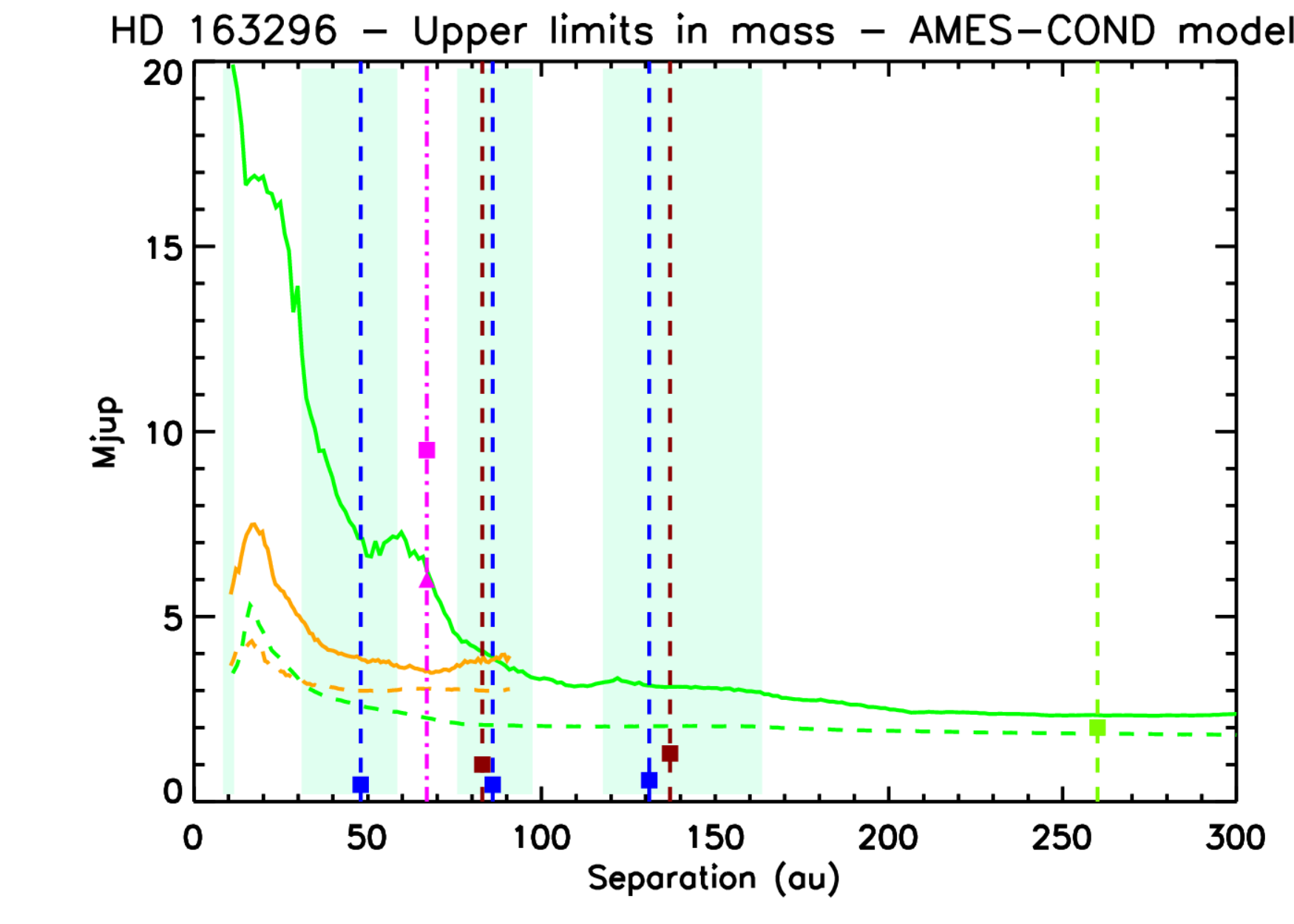}
\caption{Mass limits versus the separation from the central star expressed in
  au for IRDIS (green line) and for IFS (orange line) obtained using the
  AMES-COND models. The dashed lines in the same colours represents the mass
  limits obtained using PACO. The light cyan areas represent the estimated
  positions of the gaps around the star as estimated in
  \citet{2018ApJ...869L..49I}. We also overplot the dashed lines that
  represent the positions of the planets proposed by
  \citet{2018ApJ...857...87L}(blue lines),
  \citet{2018ApJ...860L..12T} (red lines), \citet{2018ApJ...860L..13P}
  (green line) and \citet{2018MNRAS.479.1505G} (magenta line). Further, we
  added the foreseen mass of each proposed planet denoting them with a filled
  square of the same color of the corresponding line. In the case of the planet
  proposed by \citet{2018MNRAS.479.1505G} we used a filled triangle to show
  the mass of the planet as proposed in the paper and the updated value of
  planetary mass with a filled square (see the text for more details).}
\label{f:masslimitamescond}
\end{figure}
\begin{figure}
\centering
\includegraphics[width=\columnwidth]{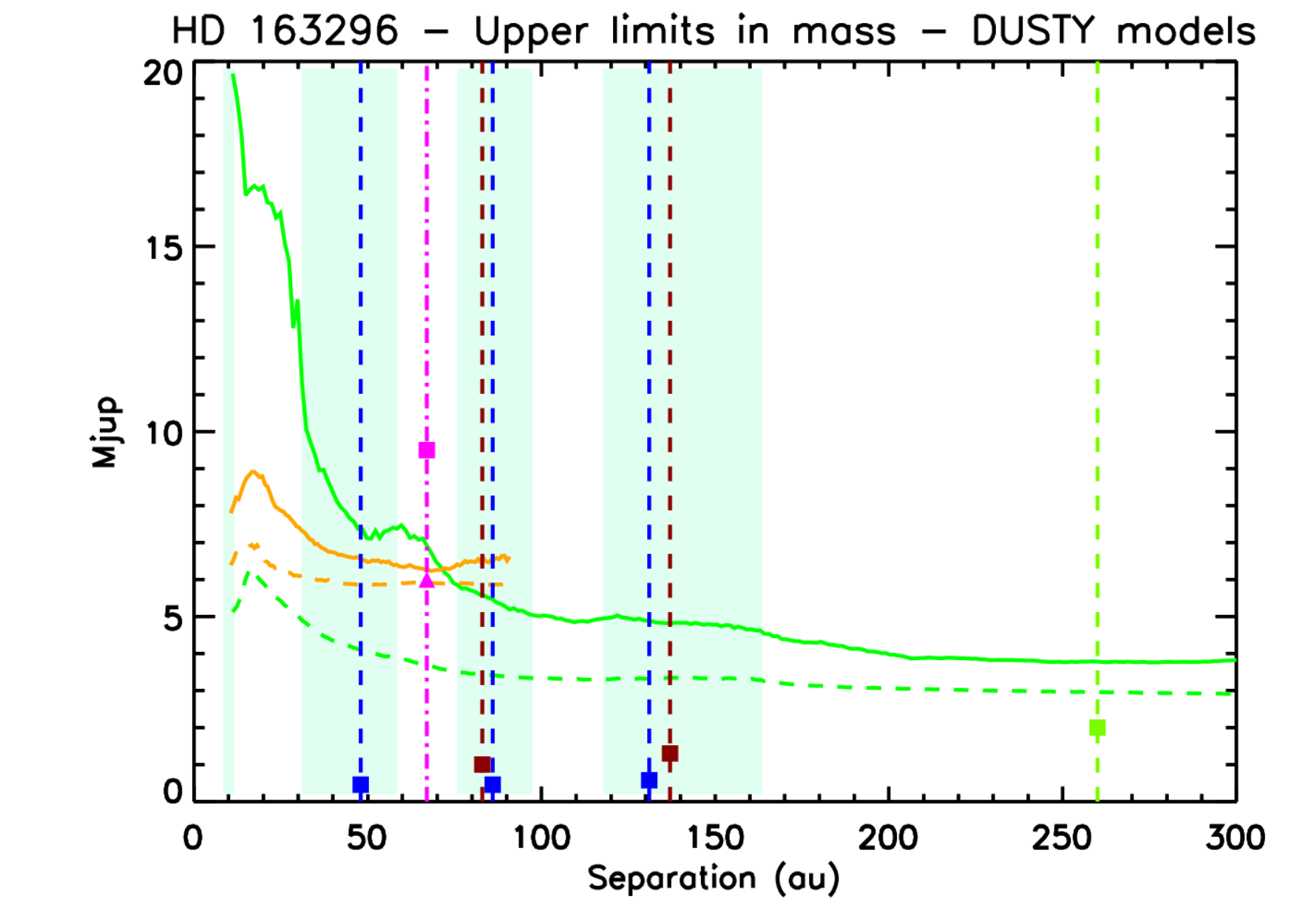}
\caption{Same than Figure~\ref{f:masslimitamescond} but for mass limits
  obtained using DUSTY models.}
\label{f:masslimitdusty}
\end{figure}

Using the contrast values obtained with the procedure described above and the
AMES-COND \citep{2003IAUS..211..325A} and the AMES-DUSTY
\citep{2001ApJ...556..357A} evolutionary models, we calculated the mass limits
for substellar objects around HD\,163296. The choice of these models gives
the possibility to explore different and extreme conditions, that are absolute
absence of clouds in the first case and complete clouds coverage in the latter
case. To this aim we assumed for the system
the age and the distance given in Section~\ref{s:intro}. Moreover, we
  assumed for the star a magnitude H of 5.53 and a magnitude K of 4.78
\citep{2003yCat.2246....0C}. The results of this
procedure are displayed in Figure~\ref{f:masslimitamescond} and in
Figure~\ref{f:masslimitdusty} where the orange lines represent the limits
obtained through IFS while the green lines represent the limits obtained through
IRDIS. As for Figure~\ref{f:contrast} we display with dashed lines the mass
limits obtained using the contrast from PACO. With PCA, IFS allows, at
separations between 30 and 80~au, to exclude the presence of sub-stellar
objects with mass larger than 3-4~\MJup if we consider the AMES-COND models
while the AMES-DUSTY models imply a larger limit between 6-7~\MJup. At larger
separations, IRDIS allows to put limits of the order
of 3-4~\MJup up to $\sim$200~au and lower than 3~\MJup at larger separations
when considering the AMES-COND models while the limits with the AMES-DUSTY
models are of 4-5~\MJup at the same separations. PACO allows to improve
especially at short separations while at larger separations the IRDIS mass
limits tends to converge to similar values obtained with the PCA method. In
the same images we colored in light cyan the zones of the gaps defined by
\citet{2018ApJ...869L..49I} and we overplotted dashed vertical lines at the
separations foreseen for the the planets proposed by
\citet{2018ApJ...857...87L}, \citet{2018ApJ...860L..12T},
\citet{2018ApJ...860L..13P} and \citet{2018MNRAS.479.1505G} adding also a
filled square indicating the mass of each proposed planet. We can use these
results to put limits at the gaps positions. The inner one, recently
discovered lay behind the SPHERE coronagraph so that we cannot put any
constraints about it. The calculated limits for the other three gaps with
both the adopted models are listed in Table~\ref{t:gaplimit} and are of the
order of a few \MJup.

It is clear, however, that the depth of our
observations is not enough to detect the planets proposed both by
\citet{2018ApJ...857...87L} and \citet{2018ApJ...860L..12T} both
considering the limits with AMES-COND and AMES-DUSTY models. On the
other hand, the planet of 6~\MJup at a projected separation of $\sim$67~au as
proposed by \citet{2018MNRAS.479.1505G} should be visible with IFS while it
should be at the detection limit with IRDIS (filled triangle in
Figure~\ref{f:masslimitamescond}) with AMES-COND models while it should be
below the detection limits with the AMES-DUSTY models. It has to be noted,
however, that the mass of 6~\MJup was obtained assuming an age of 5~Myr
obtained from \citet{2009A&A...495..901M}. If we reconsider the mass of this
companion using the age used in this work, it would have a mass between 9 and
10~\MJup well above the detection limits with both the intruments and with
both the adopted models. As written above, we are not able to detect this object
in the final images obtained both with IFS and IRDIS as enlighted in the
magenta circle displayed in Figure~\ref{f:finalimage}. To exist without
being detected in our data, this objects should be very red with a H-L$\sim$4.6
and a K-L$\sim$2.5.
Finally, the planet proposed by \citet{2018ApJ...860L..13P} should be just
below the detection limit of the IRDIS data using the AMES-COND models while
it is well below the detection limits using the AMES-DUSTY models. Probably,
given the disk environment rich of gas and dust in which the proposed planet
is located the latter would be the more adequate in this case and this might
explain why we are not able to detect it.

\begin{table}
 \centering
 \begin{minipage}{90mm}
   \caption{Mass limits expressed in \MJup at the estimated positions of
     the gaps in the disk of HD\,163296 calculated both with AMES-COND
     (third column) and AMES-DUSTY (fourth column) models. In both cases the
     limit are expressed in \MJup. In the second column we report the
     separation expressed in au of each gap as found by
     \citet{2018ApJ...869L..49I}. \label{t:gaplimit}}
  \begin{tabular}{c c c c}
  \hline
 Gap &  Separation & A.-C. M. lim.  &  A.-D. M. lim. \\
         \hline
   1 &    45  &   3.7-4.9                        &     6.4-7.3          \\
   2 &    87  &   3.4-4.5                        &     5.0-6.6          \\
   3 &   159  &   3.0-3.3                        &     4.6-5.0          \\
\hline
\end{tabular}
\end{minipage}
\end{table}

\subsection{Candidate companions}
\label{s:companion}

\begin{figure}
\centering
\includegraphics[width=\columnwidth]{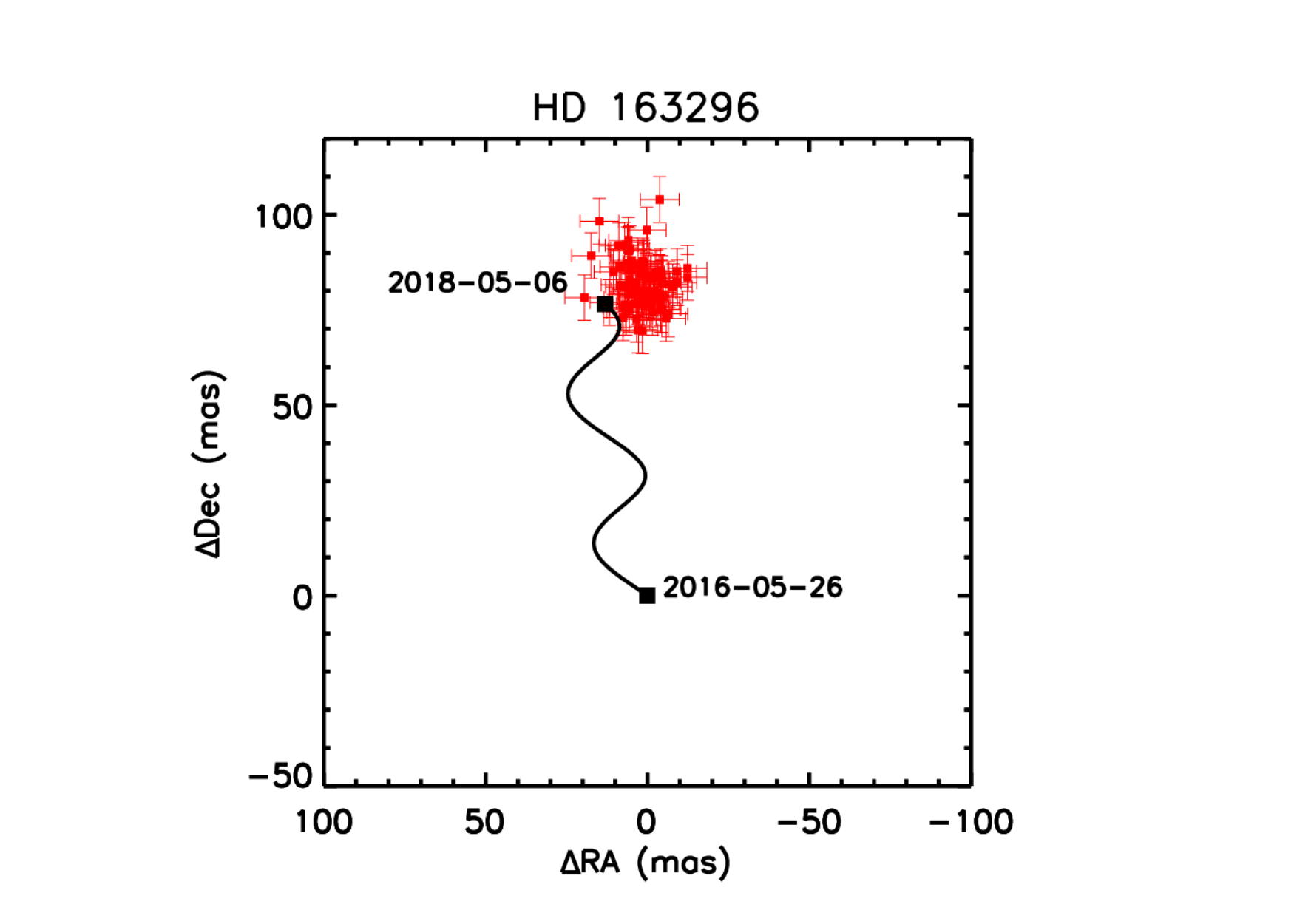}
\caption{Differential motion in right ascension and in declination of the
  cross-identified sources between the SPHERE polarized data and the last
  observing data represented by the red squares. The black line represents
  the stellar motion due to proper motion and to annual parallax. The dark
    squares indicate the differential positions, relative to the first epoch,
    that a background object would have}.
\label{f:propermotion}
\end{figure}

\begin{figure}
\centering
\includegraphics[width=\columnwidth]{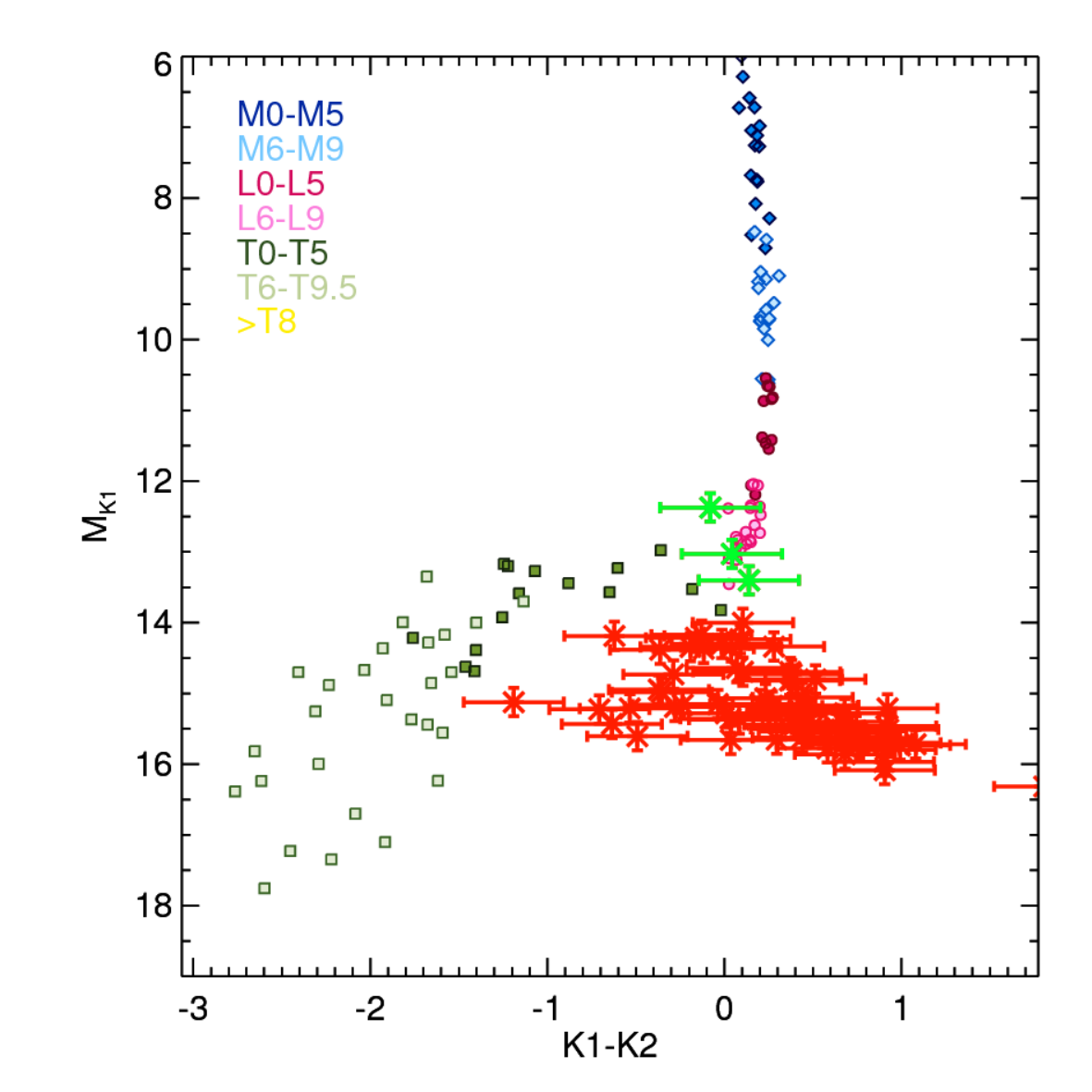}
\caption{K1 versus K1-K2 color-magnitude diagram with the positions of
  the candidates detected in the HD\,163296 IRDIS FOV for which it was not
  possible to apply the proper motion test. These objects are represented
    by red and green asterisks while squares and diamond with different
    colours represent field dwarfs with spectral types as indicated in the plot
    legend. Large part of these objects (red asterisks) are in a position of
    the diagram that confirm they are background objects. Three of them (green
    asterisks) are in a positions that does not allow a definitive conclusion
    about them.}
\label{f:cmdiagram}
\end{figure}

Not surprisingly, given that the star is in the direction of the galactic
center, the IRDIS FOV contains a lot of point-like sources. We have identified
111 of them in the first epoch and 166 in the second one when we were able to
get much deeper images. We were able to cross-identify 91 of them between
the two epochs. The objects that were identified in the first epoch but not
in the second one were all near to the edge of the IRDIS FOV in the first
epoch so that they were outside in the second one. Given the short time span
between the two epochs that did not allow a good use of the proper motion
test, we decided to use also SPHERE H-band polarized data taken in the
night 2016-05-26 and used for the work in \citet{2018A&A...614A..24M}.
In this latter case, we identify 92 point-like sources and 82 of them were
successfully cross-identified with sources in the last epoch.
All these targets are background objects as demonstrated by the proper motion
test displayed in Figure~\ref{f:propermotion}. The remaining targets
identified just in the last epoch are very low luminosity sources at
large ($>$3\as) projected separation from HD\,163296. With very high
probability they also are background objects. To further confirm this we
have plotted these objects in K1 versus K1-K2 color-magnitude diagram comparing
them with the positions of field dwarfs objects as displayed in
Figure~\ref{f:cmdiagram}. The positions of large part
of them confirm that they actually are background objects. Just for three of
them it is not possible to draw a definitive conclusion as they are in a part
of the diagram compatible with the positions of companion objects. In
Table~\ref{t:psource} we list the separations in right ascension and in
declination of the 166 point source identified in the second observing epoch
together with the status of each object. 

\begin{table*}
 \centering
 \begin{minipage}{140mm}
   \caption{List of point sources found in the HD\,163296 IRDIS FOV at
     2018-05-06 epoch. The second
     and third columns give the separation, expressed in mas, from the star
     in right ascension and in declination respectively. The status values in
     column 4 are: background object by proper motion test (1); background
     object by photometry (2); undefined (3). \label{t:psource}}
     \begin{tabular}{c c c c | c c c c | c c c c}
  \hline
    Id. & $\Delta$$\alpha$ & $\Delta$$\delta$ & Status & Id. & $\Delta$$\alpha$ & $\Delta$$\delta$ & Status & Id.  & $\Delta$$\alpha$ & $\Delta$$\delta$ & Status \\
         \hline
    1   &   -509.00  &    5462.37 &   1 & 57   &    645.59  &   -4879.25 &   1 &113   &  -5197.64  &   -3883.85 &   2\\
    2   &   -410.41  &    4994.64 &   1 & 58   &   1662.26  &    3569.96 &   1 &114   &  -4215.46  &   -3812.74 &   2\\
    3   &   -299.90  &    4538.29 &   1 & 59   &   1125.33  &   -5559.30 &   1 &115   &  -3625.25  &   -3451.66 &   2\\
    4   &   -985.98  &    4541.28 &   1 & 60   &   2319.59  &   -5044.66 &   1 &116   &  -3690.08  &   -4459.15 &   2\\
    5   &  -1676.50  &    4473.01 &   1 & 61   &   2410.89  &   -3366.61 &   1 &117   &  -2612.69  &   -4643.71 &   2\\
    6   &    270.84  &    4874.09 &   1 & 62   &   3693.00  &   -3210.14 &   1 &118   &  -2325.86  &   -4715.11 &   2\\
    7   &   1220.65  &    4940.52 &   1 & 63   &   1662.26  &    3569.96 &   1 &119   &  -2205.00  &   -3466.75 &   2\\
    8   &   2172.12  &    5120.92 &   1 & 64   &   3868.75  &   -2409.64 &   1 &120   &   -805.54  &   -3762.94 &   2\\
    9   &   2360.17  &    5030.16 &   1 & 65   &   4402.96  &   -2636.94 &   1 &121   &   -701.18  &   -3606.38 &   2\\
   10   &   2649.85  &    5399.47 &   1 & 66   &   4480.68  &   -1945.64 &   1 &122   &  -1471.83  &   -2469.98 &   2\\
   11   &   2520.20  &    4418.24 &   1 & 67   &   5136.88  &   -1493.52 &   1 &123   &   -467.72  &   -3099.78 &   3\\
   12   &   2658.56  &    4415.95 &   1 & 68   &   4771.12  &    -160.86 &   1 &124   &   -538.31  &   -4245.59 &   2\\
   13   &    529.31  &    3161.66 &   1 & 69   &   4522.26  &    -219.04 &   1 &125   &   -144.50  &   -4187.11 &   2\\
   14   &   1305.58  &    3548.11 &   1 & 70   &   3413.44  &     230.48 &   1 &126   &    255.88  &   -3992.68 &   2\\
   15   &   1401.44  &    3603.88 &   1 & 71   &   3079.69  &    -774.57 &   1 &127   &   -652.68  &   -4778.69 &   2\\
   16   &   1662.26  &    3569.96 &   1 & 72   &   3317.67  &    -649.42 &   1 &128   &     92.89  &   -4730.31 &   2\\
   17   &   1890.13  &    3473.17 &   1 & 73   &   4236.86  &    1020.44 &   1 &129   &   1226.90  &   -4587.38 &   2\\
   18   &   1569.48  &    2933.20 &   1 & 74   &   4621.78  &    1055.75 &   1 &130   &    742.31  &   -4113.75 &   2\\
   19   &   2793.61  &    3146.72 &   1 & 75   &   4740.98  &    1561.58 &   1 &131   &   1171.58  &   -5135.95 &   2\\
   20   &   3075.84  &    2745.79 &   1 & 76   &   1662.26  &    3569.96 &   1 &132   &   -489.67  &   -5879.36 &   2\\
   21   &   2510.02  &    2004.26 &   1 & 77   &   3798.52  &    2093.64 &   1 &133   &    553.83  &   -5409.67 &   2\\
   22   &   2283.82  &    1484.59 &   1 & 78   &   3973.68  &    2193.86 &   1 &134   &   1454.72  &   -5609.35 &   2\\
   23   &   2594.07  &     958.99 &   1 & 79   &   3932.87  &    2310.46 &   1 &135   &   2027.80  &   -5762.07 &   2\\
   24   &  -2258.35  &    3155.69 &   1 & 80   &   4628.56  &    2877.03 &   1 &136   &   2331.47  &   -6007.90 &   2\\
   25   &  -2338.29  &    2645.61 &   1 & 81   &   5251.38  &    2754.56 &   1 &137   &   2395.95  &   -5752.67 &   3\\
   26   &  -2726.26  &    3378.10 &   1 & 82   &   3816.38  &    3990.23 &   1 &138   &   2776.32  &   -4259.94 &   2\\
   27   &  -3496.10  &    2949.37 &   1 & 83   &   -417.92  &    4360.16 &   2 &139   &   3249.41  &   -4412.67 &   2\\
   28   &  -2881.48  &    1545.46 &   1 & 84   &   -758.96  &    4997.35 &   2 &140   &   3346.74  &   -3616.26 &   2\\
   29   &  -4463.75  &    1710.53 &   1 & 85   &  -1233.94  &    5038.10 &   2 &141   &   3114.07  &   -3905.03 &   2\\
   30   &  -4048.22  &     895.79 &   1 & 86   &  -1408.75  &    5328.75 &   2 &142   &   3899.05  &   -3396.08 &   2\\
   31   &  -4209.64  &     219.15 &   1 & 87   &  -1046.24  &    6113.48 &   2 &143   &   3254.30  &   -2404.37 &   2\\
   32   &  -4685.52  &    -297.98 &   1 & 88   &  -1696.50  &    5305.74 &   2 &144   &   3960.09  &   -1343.04 &   2\\
   33   &  -5384.59  &   -1621.05 &   1 & 89   &  -2028.89  &    5891.10 &   2 &145   &   5163.41  &     589.75 &   2\\
   34   &  -3983.37  &   -1187.56 &   1 & 90   &  -2209.40  &    5105.29 &   2 &146   &   6162.53  &     740.95 &   2\\
   35   &  -4434.98  &   -1722.93 &   1 & 91   &  -2148.81  &    3917.88 &   2 &147   &   5124.51  &    1182.32 &   2\\
   36   &  -3357.80  &   -2217.04 &   1 & 92   &  -1880.89  &    3687.65 &   2 &148   &   4756.92  &    1815.45 &   2\\
   37   &  -3020.95  &   -2852.82 &   1 & 93   &  -2309.81  &    6105.89 &   2 &149   &   3445.42  &    1555.50 &   2\\
   38   &  -2751.01  &   -2513.86 &   1 & 94   &  -3102.85  &    5124.33 &   3 &150   &   2903.25  &    2045.75 &   2\\
   39   &  -2636.81  &   -3006.37 &   1 & 95   &  -3202.02  &    4760.47 &   2 &151   &   1531.25  &    2682.75 &   2\\
   40   &  -1947.87  &   -2195.32 &   1 & 96   &  -3035.23  &    4713.71 &   2 &152   &   5546.39  &    3438.70 &   2\\
   41   &  -1663.77  &   -2166.46 &   1 & 97   &  -2986.02  &    4117.62 &   2 &153   &   4825.14  &    3080.92 &   2\\
   42   &  -1473.32  &   -1967.74 &   1 & 98   &  -2717.43  &    3279.79 &   2 &154   &   4062.91  &    3211.32 &   2\\
   43   &  -4067.00  &   -3328.25 &   1 & 99   &  -2603.90  &    2433.66 &   2 &155   &   4124.00  &    3618.58 &   2\\
   44   &  -3056.33  &   -3808.62 &   1 &100   &  -4020.27  &    2323.83 &   2 &156   &   3337.48  &    4234.68 &   2\\
   45   &  -3490.52  &   -4895.52 &   1 &101   &  -2778.04  &    1651.68 &   2 &157   &   3728.81  &    4463.81 &   2\\
   46   &  -2509.30  &   -5186.37 &   1 &102   &  -3556.73  &    1120.40 &   2 &158   &   3722.73  &    4854.00 &   2\\
   47   &  -2143.59  &   -5401.03 &   1 &103   &  -3615.01  &     941.02 &   2 &159   &   2597.00  &    3846.50 &   2\\
   48   &  -1701.97  &   -5567.30 &   1 &104   &  -4965.56  &     899.10 &   2 &160   &   2383.31  &    3969.31 &   2\\
   49   &  -1365.45  &   -5838.66 &   1 &105   &  -5122.78  &     885.34 &   2 &161   &   2484.19  &    4520.51 &   2\\
   50   &  -1297.13  &   -3768.08 &   1 &106   &  -3553.75  &     171.67 &   2 &162   &   2805.07  &    4852.63 &   2\\
   51   &   -994.06  &   -4053.32 &   1 &107   &  -5194.53  &    -278.64 &   2 &163   &   2116.53  &    3737.97 &   2\\
   52   &   -737.94  &   -4554.44 &   1 &108   &  -4223.51  &    -599.60 &   2 &164   &   2637.07  &    5139.71 &   2\\
   53   &    147.76  &   -3344.41 &   1 &109   &  -2059.83  &   -1518.15 &   2 &165   &    634.82  &    3635.36 &   2\\
   54   &   1662.26  &    3569.96 &   1 &110   &  -3883.63  &   -2274.38 &   2 &166   &    543.62  &    4228.48 &   2\\
   55   &   1139.34  &   -3052.97 &   1 &111   &  -1803.24  &   -1877.30 &   2 &   &      &     &  \\
   56   &   1306.98  &   -3703.06 &   1 &112   &  -2522.84  &   -2304.49 &   2 &   &      &     &  \\
\hline
\end{tabular}
\end{minipage}
\end{table*}

\section{Discussion and Conclusions}
\label{s:conclusion}

In this work we have presented the results of SPHERE SHINE observations of
the Herbig star HD\,163296. The effectiveness of these observations is mainly
limited by the fact that it is not possible to observe this star during the
passage through the meridian from Paranal Observatory. For this reason the
total rotation of the FOV cannot be large during the observation limiting the
contrast deepness that can be reached through high-contrast imaging techniques
like ADI. Despite this limitation we were however able to obtain a contrast
better than $10^{-5}$ at separation less than 1\as thanks to IFS and of
the order of $10^{-6}$ at separations larger than 2\as thanks to IRDIS.
This contrast allows to put mass limits between 3-4~\MJup or 6-7~\MJup at
projected separations between 30 and 80~au using the AMES-COND and AMES-DUSTY
models respectively. Furthermore, IRDIS allows to obtain mass limit of
$\sim$2~\MJup or $\sim$4~\MJup at projected separations larger than 200~au
using AMES-COND and AMES-DUSTY respectively. Given the environment around
HD\,163296 the latter is probably the more adequate to this case. The use
of both models can however give an idea of the range of variability of the
mass limit around this target. For this work we have assumed an age of the
system of 10.4~Myr obtained by a recent determination contrarily to what
was generally done in the previous works on this object that used an
older determination of the age (5~Myr). This of course results in higher
mass limits that are however more reliable than those obtained using the
younger age. \par
It is anyhow important to stress that these limits do not take into account the
effects of the material (dust or gas) of the disk on the visibility of
companions around HD\,163296. Indeed, if they are embedded in the disk or
behind it, we would expect that they are extremely reddened or suffering large
amounts of extinction so that the limits given above are valid in case of
planets with a low absorption due to the disk. Example of companions observed
with SPHERE that are embedded in the disk and, for this reason, are very
reddened or almost totally extincted are e.g. the debated companions of
HD\,100546 \citep{2018A&A...619A.160S} or the stellar companion R\,CrA\,B
\citep{2019arXiv190202536M}. There is a paucity of studies that quantify
the effect of disks on embedded planets, so that it is not possible in
this work to draw conclusions on this particular case.
In addition, we have to consider that, due to the fact that the
  HD\,163296 disk is still extremely gas-rich, planetary objects embedded into
  it would be probably surrounded by a circum-planetary disk (CPD) like
  recently proposed by \citet{2019arXiv190506370C} for the case of PDS\,70\,b.
  As demonstrated by \citet{2015ApJ...799...16Z} the flux at NIR wavelengths
  should be dominated by the disk but recently \citet{2019MNRAS.tmp.1269S}
  concluded, based on the SED, that the best contrast between the circumstellar
  disk and the CPD is for sub-mm/radio wavelengths while the CPD observation
  should be strongly hampered at NIR wavelengths.\par
Mass limits of a few \MJup are however not enough to give any conclusion
about the presence (or not) of the planetary companions proposed by
\citet{2018ApJ...857...87L} and \citet{2018ApJ...860L..12T}. On the other hand,
the companion at large separation proposed by \citet{2018ApJ...860L..13P} is
just below the detection limit obtained through IRDIS. We were however not
able to retrieve it in our data so that further observations and analysis
are mandatory to fully exclude or confirm its existence.
Finally, the mass limits that we obtain for this star should allow to
detect a planet of 6~\MJup at a projected separation of 0.49\as as proposed
by \citet{2018MNRAS.479.1505G} both with IRDIS and with IFS. The fact that
we are not able to recover this planet should rule out its presence
confirming what recently found by \citet{2019ApJ...875...38R} with
  observations in J, H and K spectral bands at the Subaru Telescope. However,
we have to remember that this planet has been discovered by observations in
the L' band. As stressed above, due to the disk environment in which it
would reside a putative planet would then experience a strong absorption. This
could induce a very red spectrum that would make difficult to image it at the
shortest wavelengths used for SPHERE. Another possible explanation could be
that this object has a very dusty and cool atmosphere. This latter case would
be very similar to those of HD\,95086 discovered in L' band with NACO by
\citet{2013ApJ...772L..15R}. This planet was then recovered with difficulty
with SPHERE using IRDIS in the K-band and only marginally in H-band with
IFS combining datasets from different epochs \citep{2018A&A...617A..76C}.
This planet could then be a similar, or even more extreme, case. \par
In conclusion, our work demonstrates that SPHERE operating in high-contrast
imaging mode is able to reach very deep contrast limits for young stellar
systems even if it is operating in not ideal conditions. While it was not
possible to give any conclusion for the lower mass companions proposed in
previous works on this system, we were able to exclude, or at least to put
strong constraints on their physical characteristics. Finally, our data seem
to exclude the presence, even if with some caveat, of the only candidate
companion until now proposed through high-contrast imaging methods. \par
High-contrast imaging instruments like SPHERE or GPI are providing state-of-art
data whose quality cannot be overcome at present. Given that their limit are
not enough to draw a definitive conclusion on the presence of the proposed
companions, we then conclude that the difficulties in confirming for this and
for similar stars that the observed disk structures are due to the interaction
with substellar objects are mainly due to technological limits. We will then
need observations with future instrumentations to be able to confirm or to
reject the hypothesis of planet/disk interaction to explain these structures.

\section*{Acknowledgments}
SPHERE is an instrument designed and built by a consortium
consisting of IPAG (Grenoble, France), MPIA (Heidelberg, Germany), LAM
(Marseille, France), LESIA (Paris, France), Laboratoire Lagrange
(Nice, France), INAF–Osservatorio di Padova (Italy), Observatoire de
Genève (Switzerland), ETH Zurich (Switzerland), NOVA (Netherlands),
ONERA (France) and ASTRON (Netherlands) in collaboration with
ESO. SPHERE was funded by ESO, with additional contributions from CNRS
(France), MPIA (Germany), INAF (Italy), FINES (Switzerland) and NOVA
(Netherlands).  SPHERE also received funding from the European
Commission Sixth and Seventh Framework Programmes as part of the
Optical Infrared Coordination Network for Astronomy (OPTICON) under
grant number RII3-Ct-2004-001566 for FP6 (2004–2008), grant number
226604 for FP7 (2009–2012) and grant number 312430 for FP7
(2013–2016). We also acknowledge financial support from the Programme National
de Plan\'{e}tologie (PNP) and the Programme National de Physique Stellaire
(PNPS) of CNRS-INSU in France. This work has also been supported by a grant from
the French Labex OSUG@2020 (Investissements d’avenir – ANR10 LABX56).
The project is supported by CNRS, by the Agence Nationale de la
Recherche (ANR-14-CE33-0018). It has also been carried out within the frame of
the National Centre for Competence in  Research PlanetS supported by the Swiss
National Science Foundation (SNSF). MRM, HMS, and SD are pleased 
to acknowledge this financial support of the SNSF. Finally, this work has made
use of the the SPHERE Data Centre, jointly operated by OSUG/IPAG (Grenoble),
PYTHEAS/LAM/CESAM (Marseille), OCA/Lagrange (Nice), Observatoire de
Paris/LESIA (Paris), and Observatoire de Lyon, also supported by a grant from
Labex  OSUG@2020 (Investissements d’avenir – ANR10 LABX56). We thank P. Delorme
and E. Lagadec (SPHERE Data Centre) for their efficient help during the data
reduction process. \par
This work has made use of data from the European Space Agency (ESA) mission
{\it Gaia} (\url{https://www.cosmos.esa.int/gaia}), processed by
the {\it Gaia} Data Processing and Analysis Consortium (DPAC,
\url{https://www.cosmos.esa.int/web/gaia/dpac/consortium}). Funding for
the DPAC has been provided by national institutions, in particular the
institutions participating in the {\it Gaia} Multilateral Agreement. \par
This research has made use of the SIMBAD database, operated at CDS,
Strasbourg, France. \par
The authors thanks Dr. G. Guidi for kindly sharing images from her work.
D.M. acknowledges support from the ESO-Government of Chile Joint Comittee
program 'Direct imaging and characterization of exoplanets'. 
D.M., A.Z., V.D.O., R.G., S.D., C.L. acknowledge support from
the ``Progetti Premiali'' funding scheme of the Italian Ministry of Education,
University, and Research. A.Z. acknowledges support from the CONICYT + PAI/
Convocatoria nacional subvenci\'on a la instalaci\'on en la academia,
convocatoria 2017 + Folio PAI77170087.
R.L. has received funding from the European Union's Horizon 2020 
research and innovation program under the Marie Sk\l odowska-Curie grant
agreement n. 664931. G.v.d.P acknowledges funding from ANR of France under
contract number ANR-16-CE31-0013 (Planet-Forming-Disks). \par

\bibliographystyle{mn2e}
\bibliography{hd163296}


\label{lastpage}

\end{document}